\documentclass[12pt]{elsart}
\usepackage{amsmath,amssymb}

\newcommand{\be}{\begin{equation}}
\newcommand{\ee}{\end{equation}}
\newcommand{\bea}{\begin{eqnarray}}
\newcommand{\eea}{\end{eqnarray}}

\newcommand{\g}{GL(n, \mathbb{R})}
\newcommand{\gt}{GL(3, \mathbb{R})}

\newcommand{\son}{SO(n, \mathbb{R})}
\newcommand{\sot}{SO(3, \mathbb{R})}

\begin{document}
\begin{frontmatter}

\title{
Generalized Calogero-Moser-Sutherland models from geodesic motion on
$GL^+(n, \mathbb{R})$ group manifold}
\author[Tbilisi,DubnaLIT]{Arsen Khvedelidze \thanksref{akhemail}}
and
\author[DubnaBLTP]{Dimitar Mladenov\thanksref{dmmemail}}

\address[Tbilisi]{
Department of Theoretical Physics,
A. Razmadze Mathematical Institute, GE-380093 Tbilisi, Georgia}

\address[DubnaLIT]{Laboratory of Information Technologies,
Joint Institute for Nuclear Research, 141980 Dubna, Russia}

\address[DubnaBLTP]{
Bogoliubov Laboratory of Theoretical Physics,
Joint Institute for Nuclear Research, 141980 Dubna, Russia}

\thanks[akhemail]{Electronic mail: khved@thsun1.jinr.ru}
\thanks[dmmemail]{Electronic mail: mladim@thsun1.jinr.ru}

\begin{abstract}
It is shown that geodesic motion on the $GL(n, \mathbb{R})$ group
manifold endowed with the bi-invariant metric
$d s^2 = \mbox{tr}\left( g^{-1} d g \right)^2$
corresponds to a
generalization of the hyperbolic $n$-particle Calogero-Moser-Sutherland model.
In particular, considering the motion on Principal orbit stratum
of the $\son$ group action, we arrive at dynamics of a generalized
$n$-particle Calogero-Moser-Sutherland system
with two types of internal degrees of freedom obeying $\son \bigoplus \son$ algebra.
For the Singular orbit strata of $\son$ group action the geodesic motion
corresponds to certain deformations of the
Calogero-Moser-Sutherland model in a sense of description of
particles with different masses.
The mass ratios depend on the type of Singular orbit stratum and are
determined by its degeneracy.
Using reduction due to discrete and continuous symmetries of the system
a relation to II$A_n$ Euler-Calogero-Moser-Sutherland model is demonstrated.
\end{abstract}

\end{frontmatter}



\section{Introduction}


Almost twenty years ago a possibility was discovered to maintain the
integrability of Calogero-Moser-Sutherland models \cite{CSM}
(classification and description can be found in \cite{PerelBook})
supposing that the particles moving on a line have additional internal
degrees of freedom \cite{GibbonsHermsen,Wojciechowski}.
Latter it was shown \cite{KBBT,BabelonTalon}
that the generic elliptic Calogero-Moser-Sutherland type system,
which consists of $n$-particles on a line interacting with pairwise
potential in the form of Weierstrass elliptic function $V(z) = \wp(z)$,
admits the following generalization
\begin{equation} \label{eq:Sgcsm}
H = \frac{1}{2} \sum_{i=1}^n p_i^2 +
\frac{1}{2} \sum_{i\not=j}^n f_{ij}f_{ji}\wp(x_i-x_j)\,.
\end{equation}
Here apart from the canonical pairs $(x_i, p_i)$,
describing the position and the momenta of particles
and obeying nonvanishing Poisson brackets
\begin{equation}
\{ x_i, p_j \} = \delta_{ij} \,,
\end{equation}
the ``internal'' degrees of freedom  $f_{ab}$ which satisfy the algebra
\begin{equation}
\{f_{ab}, f_{cd} \} =
\delta_{bc} f_{ad} - \delta_{ad} f_{cb}\,
\end{equation}
are included.
The recent comprehensive discussion of the integrability of generic
spin Calogero-Moser-Sutherland systems can be found in the papers \cite{LiXu1,LiXu2}.
One of the most effective and transparent way to convince in the integrability of a
Hamiltonian system is to find a known higher-dimensional exactly solvable model,
whose dynamics on a certain invariant submanifold coincides with the dynamics
of the given Hamiltonian system.
This method is known as symplectic reduction method
\cite{Arnold,MarsdenRatiu,BorisovMamaev}.
The Calogero-Sutherland-Moser systems
with the so-called degenerate cases of the potential,
when the Weierstrass elliptic function $\wp(z)$ reduces to
$1/\sinh^2 z$, $1/\sin^2 z$ or to the rational function $1/z^2$,
have a well-known interpretion as symplectic reductions of geodesic motions
on symmetric spaces
\cite{OlshanetskyPerelomov,KazhdanKonstantSternberg}.
Furthermore, it was argued in \cite{GorskiNekrasov} that
symplectic reduction relates the
elliptic Calogero-Moser-Sutherland system with
certain integrable Hamiltonian system on the
cotangent bundle to the central extension of two-dimensional Lie algebra of
$SL(n, \mathbb{C})$-valued currents on the some elliptic curve.
New types of generalizations of the spin Calogero-Moser-Sutherland
systems with nonstandard spin interactions
have been constructed in \cite{PolyDiscrete}
using discrete symmetries of the model.

In the present Letter we shall exploit the idea of symplectic
reduction considering certain generalization of the
Calogero-Sutherland-Moser model.
Namely, we shall consider the integrable finite-dimensional model
corresponding to the geodesic motion on the general linear matrix group with a positive
determinant $GL^+(n, \mathbb{R})$,\footnote{
Hereafter we shall omit the upper index $+$
to simplify the expressions.}
endowed with the left- and right-invariant metric
$d s^2 = \mbox{tr}\left( g^{-1} d g \right)^2$, where $g \in \g$.
In terms of this bi-invariant metric on $\g$ group manifold
the equations of motion for the corresponding dynamical system
are encoded in the Lagrangian
\cite{Arnold,MarsdenRatiu}
\begin{equation}\label{eq:bilag}
L_{GL} = \frac{1}{2}\, \mbox{tr}\left( g^{-1} \dot g \right)^2\,,
\end{equation}
where overdot denotes differentiation with respect to time.
Bellow we shall represent the Hamiltonian corresponding to
Lagrangian (\ref{eq:bilag}) in terms of a special
parameterization, adapted to the action of $\son$ symmetry
group of the system.
We shall demonstrate that on the Principal
orbit stratum of $\son$ group action the resulting Hamiltonian defines a new
generalization of the Calogero-Sutherland-Moser model
by introducing two internal variables ``spin'' and ``isospin''.
Furthermore, performing the Hamiltonian reduction owing to two types of symmetry:
continuous and discrete, we show
how to arrive at the conventional Hamiltonian of
Euler-Calogero-Sutherland model
\begin{equation} \label{eq:ECS}
H = \frac{1}{2} \sum_{i=1}^n p_i^2 +
\frac{1}{8} \sum_{i \not=j}^n \frac{l^2_{ij}}{\sinh^2(x_i-x_j)}
\end{equation}
with internal variables $l_{ab} = - \, l_{ba}$, obeying the $\son$
Poisson bracket algebra
\begin{equation}
\{l_{ab}, l_{cd} \} =
\delta_{ac} l_{bd} - \delta_{ad} l_{bc} +
\delta_{bd} l_{ac} - \delta_{bc} l_{ad}
\,.
\end{equation}

Another interesting systems arise when the dynamics takes place on
the Singular orbit strata of the $\son$ group action.
We found in this case new models representing
a certain class of mass deformed Calogero-Moser-Sutherland models.
In particular, for the case of $\gt$ group, our analysis shows that the dynamics on
Singular orbit stratum with isotropy group $SO(2)\otimes Z_2$,
corresponds to Calogero-Moser-Sutherland model, describing two particles,
whose mass ratios is $m_1 : m_2 = 2 : 1$
(see eq. (\ref{eq:HSOr2})).
The question of integrability of the mass deformed
Calogero-Moser-Sutherland models has been discussed in
\cite{Veselov} and references therein.


\section{Geodesic motion on the Principal orbit stratum}


\subsection{Symmetries and dynamics}\label{sec:PO}

If we choose the elements of the matrix $g \in \g$ as $n^2$ Lagrangian coordinates,
the Euler-Lagrange equations obtained from the Lagrangian (\ref{eq:bilag})
can be represented in the form of current conservation
\begin{equation}\label{eq:intl}
\frac{d}{dt}\left(g^{-1} \dot g \right)\, =\, 0\,.
\end{equation}
This form allows to find the general solution, i.e. the geodesics of the
bi-invariant metric are given by
\begin{equation}
g(t) = g(0) \, \exp{(tJ)}\,,
\end{equation}
where $g(0)$ and $J$ are two arbitrary constant matrices.
The special choice of these matrices corresponds to
the particular solutions describing the motion on a certain invariant
submanifold.

To show that (\ref{eq:intl}) are equations of motion of many-particle
system representing a certain generalization of the Calogero-Moser-Sutherland model,
it is useful to pass to the Hamiltonian form
of the geodesic motion on $\g$ group.
Performing Legendre transformation of the Lagrangian (\ref{eq:bilag})
\begin{equation}
\pi^T_{ab}\, = \, \frac{\partial \, L_{GL}}{\partial\, \dot g_{ab}} =
\left( g^{-1}\, \dot g\, g^{-1} \right)_{ab}
\end{equation}
we arrive at the canonical Hamiltonian
\begin{equation}\label{eq:biham}
H_{GL} = \frac{1}{2}\, \mbox{tr} \left( \pi^T  g \right)^2
\end{equation}
generating the Hamilton equations of motion
\begin{eqnarray}
\label{eq:hameq1}
&& \dot g = \{ g \,, H_{GL}\}\, =\, g \, \dot \pi^T \, g \,, \\
\label{eq:hameq2}
&& \dot\pi = \{ \pi\,, H_{GL}\}\, =\,  - \pi \,g^T\, \pi\,.
\end{eqnarray}
The nonvanishing Poisson brackets between the fundamental phase space variables
$(g_{ab}, \pi_{ab})$ are
\begin{equation}
\{g_{ab} \,, \pi_{cd}\} = \delta_{ac}\, \delta_{bd}\,.
\end{equation}

From now on the purpose of the present paper will be
to rewrite this Hamiltonian in terms of coordinates,
adapted to the symmetry possessing the system.
At first we would like to analyze the following symmetry action of the
$\son$ group on $\g$
\begin{equation}\label{eq:gact}
g \mapsto  g^\prime = R \, g\,
\end{equation}
with time-independent orthogonal matrix $R$.
In order to consider the configuration space as manifold with orbit and slice
structure with respect to this action, it is convenient to use the polar decomposition
\cite{Zelobenko} for an arbitrary element of the $\g$ group.
For the sake of technical simplicity we investigate in details
the $\gt$ group hereinafter, i.e.
\begin{equation}\label{eq:polar}
g = O S\,,
\end{equation}
where $S$ is a positive definite $3 \times 3$ symmetric matrix, and
$O(\phi_1,\phi_2, \phi_3) = e^{\phi_1 J_3}e^{\phi_2 J_1}e^{\phi_3 J_3}$
is an orthogonal matrix with $SO(3, \mathbb{R})$ generators in
adjoint representation $(J_a)_{ij} = \varepsilon_{iaj}$.
Since the matrix $g$ represents an element of $\gt$ group,
we can treat the polar decomposition (\ref{eq:polar}) as a
uniquely invertible transformation from the configuration variables $g$
to a new set of Lagrangian variables: six coordinates $S_{ij}$
and three coordinates $\phi_i$.
In terms of these new variables the Lagrangian (\ref{eq:bilag})
can be rewritten as
\begin{equation}\label{eq:LagPolar}
L_{GL} = \frac{1}{2}\,
\mbox{tr} \, \left(
\Theta_L + \dot S \, S^{- 1}
\right)^2 \,,
\end{equation}
where $\Theta_L := O^{-1}\, \dot O$ is a left-invariant 1-form on the
$\sot$ group.
To find the corresponding Hamiltonian
we note that the polar decomposition (\ref{eq:polar}) induces the point
canonical transformation from variables $(g_{ab}, \pi_{ab})$
to new canonical pairs
$(S_{ab}, P_{ab})$ and $(\phi_a, P_a)$ obeying the nonvanisning
Poisson bracket relations
\begin{eqnarray}
&&\{ S_{ab} \,, P_{cd} \} =
\frac{1}{2}
\left( \delta_{ac}\, \delta_{bd} + \delta_{ad} \, \delta_{bc} \right)\,,\\
&&\{ \phi_{a} \,, P_{b} \} =  \delta_{ab}\,.
\end{eqnarray}
The expression of the old $\pi_{ab}$ as a function of the new coordinates is
\begin{equation}
\pi = O \left( P - k_a J_a \right)\,,
\end{equation}
where
\begin{equation}
k_a = \gamma^{-1}_{ab} \left(\eta^L_b -
\varepsilon_{bmn}\left(S P \right)_{mn}  \right)\,,
\end{equation}
$
\gamma_{ik} = S_{ik} - \delta_{ik}\, \mbox{tr}  S
$
and $\eta^L_a$ are three left-invariant vector fields on $\sot$ group
\begin{eqnarray}
&&\label{eq:liv1}
\eta^L_1 =
\frac{ \sin\phi_3 }{\sin\phi_2 }\,  P_1 +
\cos\phi_3 \,  P_2 -
\cot\phi_2 \sin\phi_3 \  P_3 \,, \\
&&\label{eq:liv2}
\eta^L_2 =
\frac{ \cos\phi_3 }{ \sin\phi_2 }\,  P_1 -
\sin\phi_3 \,  P_2 -
\cot\phi_2 \cos\phi_3 \ P_3 \,,\\
&&\label{eq:liv3}
\eta^L_3 = P_3\,.
\end{eqnarray}

Hence, in terms of the new variables, the canonical Hamiltonian (\ref{eq:biham})
takes the form
\begin{equation} \label{eq:hams}
H_{GL} =
\frac{1}{2}\, \mbox{tr} \left( PS \right)^2 +
\frac{1}{2}\, \mbox{tr} \left( J_a S J_b S \right) k_a k_b\,,
\end{equation}
where the canonical variables $(S_{ab}, P_{ab})$  are invariant under the transformation
(\ref{eq:gact}), while the angular variables
$(\phi_a, P_a)$ undergo changes generating by the
right-invariant Killing vector fields $\eta^R_a$
\begin{eqnarray}
&&\label{eq:Ofields1}
\eta^R_1 =
 - \sin\phi_1 \cot\phi_2 \ P_1 +
\cos\phi_1 \,  P_2 +
\frac{\sin\phi_1}{\sin\phi_2}\, P_3 \,,\\
&&\label{eq:Ofields2}
\eta^R_2 =
\,\,\cos\phi_1 \cot\phi_2 \, P_1 +
\sin\phi_1 \,  P_2 -
\frac{\cos\phi_1}{\sin\phi_2}\, P_3 \,, \\
&&\label{eq:Ofields3}
\eta^R_3 = P_1\,,
\end{eqnarray}
whose Poisson brackets with the left-invariant vector fields $\eta^L_a$ vanish
$\{\eta^L_a \,, \eta^R_b\} = 0$.

Now we pass to analysis of another type of symmetry action of the
orthogonal group.
The Lagrangian (\ref{eq:bilag}) is invariant under the
transformations
\begin{equation}\label{SO3action}
g \mapsto g^\prime  = R^T \, g\, R\,
\end{equation}
with constant orthogonal matrix $R \in \sot$.
After implementation of the polar decomposition the symmetry
transformation reads
\begin{equation}\label{SO3actionm}
S^\prime = R^T \, S\, R\,, \qquad O^\prime = R^T \, O\, R\,.
\end{equation}
The orbit space of the action $S \mapsto R^T \, S\, R$
of the $\sot$ group in the space of $3 \times 3$ symmetric matrices $\mathcal S$
is given as a quotient ${\mathcal S}/\sot$.
The quotient space ${\mathcal S}/\sot$ is a stratified manifold;
orbits with the same isotropy group are collected into {\it strata}
and uniquely parameterized by the
set of ordered eigenvalues of the matrix  $S\ $: $x_1\leq x_2 \leq x_3$.
The strata are classified according to the isotropy groups
which are determined by the degeneracies of the matrix eigenvalues:
\begin{enumerate}
\item
{\it Principal orbit-type stratum},
when all eigenvalues are unequal $x_1< x_2 < x_3$,
with the smallest isotropy group $Z_2\otimes Z_2$\,.
\item
{\it Singular orbit-type strata}
forming the boundaries of the orbit space with
\begin{enumerate}
\item two coinciding eigenvalues
(e.g. $x_1 = x_2$), when the isotropy group is $SO(2)\otimes Z_2$\,.
\item all three eigenvalues are equal
($x_1 = x_2 = x_3$),
here the isotropy group coincides with the isometry group $\sot$.
\end{enumerate}
\end{enumerate}

To write down the Hamiltonian describing the motion on the Principal orbit stratum,
we introduce coordinates along the slices $x$ and along the orbits $\chi$.
Namely, since the matrix $S$ is positive definite and symmetric,
we use the main-axes decomposition in the form
\begin{equation}\label{eq:mainaxes}
S = R^T(\chi) \, e^{2X} \, R(\chi)\,,
\end{equation}
where $R(\chi) \in \sot$ is an orthogonal matrix parameterized
by three Euler angles $\chi = (\chi_1, \chi_2, \chi_3)$,
and the matrix $e^{2X}$ is diagonal
$e^{2X} = \mbox{diag}\, \| e^{2x_1}, e^{2x_2}, e^{2x_3} \|$.
The momenta $p_i$ and $p_{\chi_i}$,
canonically conjugated to the eigenvalues $x_i$ and
the angles ${\chi_i}$ correspondingly,
\begin{equation}
\{x_i \,, p_j \} =
\delta_{ij}\, \,,\qquad
\{ \chi_{i} \,, p_{\chi_j}\} = \delta_{ij} \,,
\end{equation}
can be found using the condition of canonical invariance of the
symplectic 1-form
\begin{equation}
\sum^3_{i,j = 1} P_{ij}\,\dot{S}_{ij}\, dt  =
\sum^3_{i = 1}   p_i \, \dot x_i \, dt  +
\sum^3_{i = 1}   p_{\chi_i}\, \dot{\chi}_i\, dt\,.
\end{equation}
The original momenta $P_{ij}$ are expressed in terms of the new canonical
pairs $(x_i, p_i)$ and $({\chi_i}, p_{\chi_i})$ as
\begin{equation} \label{eq:newmomenta}
P = R^T e^{- X}
\left(
\sum_{a=1}^{3}{\bar{\mathcal P}}_a {\bar\alpha}_a +
\sum_{a=1}^{3}{\mathcal P}_a {\alpha}_a
\right) e^{-X} R\,,
\end{equation}
with
\begin{eqnarray}
&& {\bar{\mathcal P}}_a = \frac{1}{2} \, p_a \,, \\
&& {\mathcal P}_a  = -  \frac{\xi^R_a}{4 \, \sinh(x_b - x_c)}\,,
\qquad
(\mbox{cyclic}\,\,\,\, \mbox{permutation} \,\,\, a\not=b\not= c)\,.
\end{eqnarray}
In the representation (\ref{eq:newmomenta}), we introduced the orthogonal basis
for the symmetric $3 \times 3$ matrices
$ \alpha_A = ( \overline{\alpha}_a ,\ \alpha_a ), \, a = 1, 2, 3 $
with the scalar product
\begin{equation}
\mbox{tr} (\bar\alpha_a\, \bar\alpha_b) = \delta_{ab}\,,
\quad
\mbox{tr} (\alpha_a\, \alpha_b) = 2\delta_{ab}\,,
\quad
\mbox{tr} (\bar\alpha_a\, \alpha_b) = 0
\end{equation}
and the $\sot$ right-invariant Killing vectors
\begin{eqnarray}
&& \label{eq:rf1}
\xi^R_1 =
 - \sin\chi_1 \cot\chi_2 \ p_{\chi_1} +
\cos\chi_1 \  p_{\chi_2} +
\frac{\sin\chi_1}{\sin\chi_2}\ p_{\chi_3} \,,\\
&& \label{eq:rf2}
\xi^R_2 =
\,\,\cos\chi_1 \cot\chi_2 \ p_{\chi_1} +
\sin\chi_1 \  p_{\chi_2} -
\frac{\cos\chi_1}{\sin\chi_2}\ p_{\chi_3} \,, \\
&& \label{eq:rf3}
\xi^R_3 =  p_{\chi_1}\,.
\end{eqnarray}
Thus, after passing to main-axes variables
$(x_i, p_i)$ and $({\chi_i}, p_{\chi_i})$,
the canonical Hamiltonian reads
\begin{equation} \label{eq:hamsma}
H_{GL} =
\frac{1}{8} \sum_{a=1}^{3} p_a^2 +
\frac{1}{16} \sum_{(abc)}  \frac{(\xi^R_a)^2}{\sinh^2(x_b - x_c)} -
\frac{1}{4} \sum_{(abc)}
\frac{\left(R_{am} \eta^L_m + \frac{1}{2}\, \xi^R_a \right)^2}{\cosh^2(x_b - x_c)}
\,.
\end{equation}
Here $(abc)$ means cyclic permutations $a \neq b \neq c$.
Hence we conclude that the integrable dynamical system describing a free motion on the
Principal orbit stratum can be interpreted in the adapted basis, as Generalized
Euler-Calogero-Moser-Sutherland model.
The generalization consists in the introduction of two types of internal
dynamical variables $\xi$ and $\eta$ --- ``spin'' and ``isospin'' degrees of freedom.
From their explicit expressions
(see eqs. (\ref{eq:liv1})-(\ref{eq:liv3}) and (\ref{eq:rf1})-(\ref{eq:rf3}))
it follows that they satisfy $\sot \bigoplus \sot$ Poisson bracket algebra
\begin{eqnarray}\label{LeftRight}
&& \{ \eta^L_a \,, \eta^L_b \} =  - \varepsilon_{abc} \eta^L_c \,, \\
&& \{ \xi^R_a  \,, \xi^R_b  \} =    \varepsilon_{abc} \xi^R_c \,,  \\
&& \{ \eta^L_a  \,, \xi^R_b \} = 0 \,.
\end{eqnarray}


\subsection{Lax-pair for the Generalized Euler-Calogero-Moser-Sutherland model}
\label{sec:laxpair}


In order to find a Lax representation for the generalized
Euler-Calogero-Moser-Sutherland model (\ref{eq:hamsma}) let us
consider the integrals of the geodesic motion on the Principal orbit sratum.
The integrals of motion can be written in Hamiltonian form,
following from (\ref{eq:intl}), as
\begin{equation}
J_{ab} = (\pi^T g)_{ab}\,.
\end{equation}
The algebra of this integrals realizes on the symplectic level the
$\g$ algebra
\begin{equation}
\{J_{ab}, J_{cd}\} = \delta_{bc} J_{ad} - \delta_{ad} J_{cb}\,.
\end{equation}
After the transformation to scalar and rotational variables
(\ref{eq:mainaxes}), the expression for the current $J$ reads
\begin{equation}\label{eq:intr}
J = \frac{1}{2} \sum_{a=1}^3  R^T
 \left(p_a {\bar\alpha}_a - i_a \alpha_a - j_a J_a
\right)R\,,
\end{equation}
where
\begin{equation}\label{eq:int+-}
i_a = \sum_{(abc)} \frac{1}{2}\,\,\xi^R_a \coth(x_b - x_c) +
\left(R_{am}\eta^L_m + \frac{1}{2}\,\, \xi^R_a \right) \tanh(x_b - x_c)
\end{equation}
and
\begin{equation}
j_a = R_{am}\eta^L_m + \xi^R_a \,.
\end{equation}

Using expressions (\ref{eq:intr}) for the integrals $J_{ab}$
the classical equations of motion for
Generalized Euler-Calogero-Moser-Sutherland model
can be rewritten in the Lax
form\footnote{
We set here aside the constructions of the Lax pairs with a spectral parameter.
The  Lax representations with a spectral parameter for the spin
Calogero-Moser-Sutherland models associated with the root systems
of simple Lie algebras were constructed in
\cite{LiXu1,LiXu2,DPLaxRep}.}
\begin{equation}
\dot{L} = [A, L ]\,,
\end{equation}
where the $3 \times 3$ matrices are given explicitly as

\begin{equation}\label{eq:L-Amatrices}
L = \left(
\begin{array}{ccc}
p_1       &  L^+_3,  &     L^-_2  \\
L^-_3,    &   p_2    &     L^+_1  \\
L^+_2,    &  L^-_1,  &     p_3    \\
\end{array}
\right)\,, \quad\quad \quad A = \frac{1}{4}\left(
\begin{array}{ccc}
   0      & - A_3,     &   A_2  \\
  A_3,    &    0,      & - A_1  \\
- A_2,    &   A_1,     &    0   \\
\end{array}
\right) \,.
\end{equation}
Entries $A_a$ and $L^\pm_a$ of the matrices (\ref{eq:L-Amatrices})
are given as
\begin{eqnarray}
&& L^\pm_a = - \frac{1}{2}\,\,\xi^R_a \coth(x_b - x_c) -
\left(R_{am} \eta^L_m + \frac{1}{2}\,\,\xi^R_a \right)\tanh(x_b - x_c) \pm
\left(R_{am} \eta^L_m + \xi^R_a  \right)
\end{eqnarray}
and
\begin{eqnarray}
&& A_a = \frac{\xi^R_a}{ 2 \, \sinh^2 (x_b - x_c)} -
\frac{ R_{am} \eta^L_m + \frac{1}{2}\,\,\xi^R_a }{ \cosh^2 (x_b - x_c) } \,,
\end{eqnarray}
where $(a,b,c)$ means cyclic permutations of $(1,2,3)$.

Below relations to the standard Euler-Calogero-Moser-Sutherland model
(\ref{eq:ECS}) will be demonstrated.


\subsection{Reduction to Euler-Calogero-Moser-Sutherland model}



\subsubsection{Reduction using discrete symmetries}


Now we shall demonstrate how the II$A_3$ Euler-Calogero-Moser-Sutherland model
arises from the canonical Hamiltonian (\ref{eq:biham}) after projection onto a
certain invariant submanifold determined by discrete symmetries.
Let us impose the condition of symmetry of the matrices $g \in \gt$
\begin{equation}\label{eq:primconst}
\psi^{(1)} = g - g^T = 0 \,.
\end{equation}
In order to find an invariant submanifold, it is necessary to supplement the constraints
(\ref{eq:primconst}) with the new ones
\begin{equation}\label{eq:secconst}
\psi^{(2)} = \pi - \pi^T = 0 \,.
\end{equation}
Indeed, using the Hamilton equations (\ref{eq:hameq1}) and (\ref{eq:hameq2}),
one can check that the surface defined by the set of constraints
$\Psi_A = ( \psi^{(1)}, \psi^{(2)})$
represents an invariant submanifold in the $\gt$ phase space
\begin{equation}
\dot \psi^{(1)}\vert_{\Psi_A = 0} = 0\,, \qquad
\dot \psi^{(2)}\vert_{\Psi_A = 0} = 0
\end{equation}
and the dynamics of the corresponding induced system is governed by the
reduced Hamiltonian
\begin{equation}\label{eq:pr}
H_{\gt}\vert_{\Psi_A = 0} =
\frac{1}{2} \, \mbox{tr} \left( \pi g \right)^2\,.
\end{equation}
The matrices $g$ and $\pi$ are now symmetric nondegenerate matrices,
and one can be convinced that this expression leads to the
Hamiltonian of the II$A_3$ Euler-Calogero-Moser-Sutherland model.
To prove this, it is necessary to note that the Poisson matrix
$ C_{AB} = \| \{ \psi^{(1)}, \psi^{(2)}\}\| $ is not degenerate
and after projection on the invariant submanifold, the canonical
Poisson structure is changed according to the Dirac prescription
\begin{equation} \label{eq:Dirbr}
\{F, G\}_D = \{F, G\}_{PB} - \{F , \psi_A \} C_{AB}^{-1} \{\psi_B, G\}
\end{equation}
for arbitrary functions $F$ and $G$.
The resulting fundamental Dirac brackets are
\begin{equation}
\{ g_{ab} \,, \pi_{cd} \}_D =
\frac{1}{2}\,
\left( \delta_{ac}\, \delta_{bd} + \delta_{ad} \, \delta_{bc} \right)\,.
\end{equation}
If now we introduce as in Section (\ref{sec:PO})
the main-axes variables $(x_a, \, p_a)$ and $(\chi_a, \, p_{\chi_a})$
instead of the symmetric variables $(g_{ab} \,, \pi_{ab})$
and use the representations (\ref{eq:mainaxes}) and (\ref{eq:newmomenta}),
then one can convince that the projected
Hamiltonian (\ref{eq:pr}), up to the time rescaling $t \mapsto 4t$,
governs the same dynamics as the Hamiltonian (\ref{eq:ECS}) of
II$A_3$ Euler-Calogero-Moser-Sutherland model with the intrinsic spin
variables $l_{ij} = \varepsilon_{ijk}\xi^R_k $.


\subsubsection{Reduction due to continuous symmetry}


Let us now derive a reduced Hamiltonian system employing
certain continuous symmetries of the model.
For the Hamiltonian
(\ref{eq:biham}) all angular variables are gathered in the three
left-invariant vector fields $\eta^L_a$ and thus the
corresponding right-invariant fields $\eta^R_a$
(\ref{eq:Ofields1})-(\ref{eq:Ofields3})
are integrals of
motion
\begin{equation}
\{\eta^R_a, H_{GL}\} = 0\,.
\end{equation}
The surface in the phase space determined by the constraints
\begin{equation} \label{con1}
\eta^R_a = 0
\end{equation}
defines an invariant submanifold.
These constraints obey the algebra
$ \{ \eta^R_a, \eta^R_b\} = \varepsilon_{abc}\eta^R_c $
and according to the Dirac terminology \cite{DiracL,HenTeit}
are first class constraints, this means that
after projection on the constraint shell (\ref{con1}),
the corresponding cyclic coordinates disappear
from the projected Hamiltonian.
To prove this one can use the
relation $ \eta^R_a = O_{ab}\eta^L_b $ between the left and the
right-invariant Killing vector fields.
Then, after projection to the constraint surface (\ref{con1}), the Hamiltonian
(\ref{eq:hamsma}) reduces to
\begin{equation}\label{eq:esham+}
H_{GL}\vert_{\eta^R_a = 0} =
\frac{1}{8} \sum_a^3 p_a^2 +
\frac{1}{4} \sum_{(abc)}  \frac{(\xi^R_a)^2}{\sinh^2 2(x_b - x_c)}
\,.
\end{equation}
After rescaling of the variables $2x_a \mapsto x_a$, one can be convinced
that the derived Hamiltonian coincides with the
Euler-Calogero-Moser-Sutherland Hamiltonian (\ref{eq:ECS}), where
the intrinsic spin variables are
$l_{ij} = \varepsilon_{ijk}\xi^R_k$.

As it was outlined in Section (\ref{sec:laxpair}) apart from the
integrals $\eta^R_a$ the system (\ref{eq:biham}) possesses the
integrals (\ref{eq:intr}).
Using these integrals one can choose
different invariant submanifold and to derive the corresponding
reduced system.
Here we would like only to mention that after
performing reduction to the surface defined by the vanishing
integrals $j_a = 0$, we again arrive at the
Euler-Calogero-Moser-Sutherland system.


\section{Geodesic motion on the Singular orbit strata}


In the previous sections we have investigated the geodesic motion on the
Principal orbit stratum, i.e. under the supposition that the
symmetric matrix $S$ in the polar representation (\ref{eq:polar})
has three different eigenvalues.
We now turn our attention to dynamical system corresponding
to the geodesic motion on the Singular orbit strata.
For the sake of technical simplicity, we restrict
ourselves to invariant submanifold of the phase space, defined
by $\eta^R = 0$ and consider a geodesic motion on the Singular
orbit stratum with two coinciding eigenvalues of the matrix $S$ in the
case of $\gt$ group.
Below we use two alternative methods.
At first a special parameterizations of the subspace of $3 \times 3$ symmetric
matrices with two coinciding eigenvalues  are exploited and as a result we find
that the Hamiltonian system describing the
geodesic motion on the $4$-dimensional Singular orbit stratum
is II$A_2$ Calogero-Moser-Sutherland model with particle mass ratio
$m_1 : m_2 = 2 : 1$.
Afterwards, based on the observation that the Singular orbits of
the configuration space represent the boundary of Principle orbit,
using appropriate limiting procedure
we derive from the Hamiltonian (\ref{eq:hamsma})
again a mass deformed II$A_2$ Calogero-Moser-Sutherland model
with the same particle mass ratio $m_1 : m_2 = 2 : 1$.


\subsection{
Mass deformed Calogero-Moser-Sutherland model via explicit
parameterizations of the Singular orbit stratum}
\label{sec:level3}


The Singular orbits have continuous isotropy groups and this leads to the
modification of geodesic motion.
For the case we are inteesting in,  $\gt$ group and two equal eigenvalues
of the symmetric matrix $S$, it is $SO(2)\otimes Z_2$.

The linear space of the real symmetric $n \times n $
matrices with two coinciding eigenvalues
has a real dimension \cite{Mehta}
\begin{equation}\label{eq:dim}
\mbox{dim}\, S(n) - \mbox{dim}\, S(2) + 1.
\end{equation}
Hence, we are able to parameterize such a subspace of $\gt$ group by four
real independent parameters
\begin{equation}\label{eq:singS}
S_{ab} = e^{2 x} \, \delta_{ab} - 2 \, e^{x + y}\,\sinh( x - y)\, n_a n_b \,,
\end{equation}
where $n_a$ is a unit 3-dimensional vector
\begin{equation}
n_a = \left( \sin\theta\,\sin\phi, \, \sin\theta\, \cos\phi, \, \cos\theta \right).
\end{equation}

We infer from the expression for the bi-invariant metric on the $\g$ group
that the metric induced on the $4$-dimensional Singular orbit stratum
parameterizing according to (\ref{eq:singS}) is
\begin{equation}\label{IndMetric}
\mbox{tr} \left( S^{-1} \, d S \right)^2 =
8 \, d x^2 + 4 \, d y^2 + 8 \, \sinh^2(x - y)
\left( d \theta^2 + \sin\theta\, d \phi^2
\right)\,.
\end{equation}
Therefore the Lagrangian
$ L = \frac{1}{2} \, \mbox{tr} \left(S^{-1} \dot S \right)^2 $
on the Singular orbit stratum can be written as
\begin{equation}\label{eq:SingLag}
L = 4 \, \dot x^2 +   2\, \dot y^2 + 4 \,\sinh^2 (x - y)\, \dot
n^2 \,,
\end{equation}
where
\begin{equation}
\dot n^2 = \dot\theta^2 + \sin^2 \theta \, \dot \phi^2 \,.
\end{equation}
The Legendre transformation gives the canonical Hamiltonian
\begin{equation}\label{eq:HSOr2}
H^{(2)}_{\gt} =
\frac{1}{16} p_x^2 + \frac{1}{8}\, p_y^2 +
\frac{l^2}{16 \, \sinh^2(x - y)}\,,
\end{equation}
where
\begin{equation}
l^2 = p^2_\theta + \frac{p^2_\phi}{\sin^2 \theta}  \,.
\end{equation}
Now taking into account that $l^2$ is a constant of motion we
convince that the geodesic motion with respect to the bi-invariant
metric on the $4$-dimensional Singular orbit stratum corresponds to 2-particle
mass-deformed Calogero-Moser-Sutherland model with particle mass ratio
$m_1 : m_2 = 2 : 1$.
Following this interpretation of the Hamiltonian system (\ref{eq:HSOr2})
in terms of particles,
one can say that the motion on this Singular orbit stratum corresponds to some
``gluing'' of two particles and formation of a bound particle with double mass.
It is apparent that using the center-mass coordinates the obtained
4-dimensional Hamiltonian system (\ref{eq:HSOr2})
can be reduced to 2-dimensional integrable model.


\subsection{
Mass deformed Calogero-Moser-Sutherland model via limiting procedure
from the free motion on the Principal orbit stratum
}


As it was mentioned above the Singular orbits with two coinciding eigenvalues
form the boundary of the Principle orbit stratum.
Based on this observation, now we would like to extend the Hamiltonian (\ref{eq:hamsma}),
given on the Principal orbit startum, to its boundary
by introducing the constraints that force the dynamics specially
to the neighborhood of the boundary and then use a
limiting procedure.
Since the Hamiltonian (\ref{eq:hamsma}) has a singularities along the
Singular orbits, we impose the following constraints on the phase space variables
\begin{equation}\label{eq:EpsConst}
\chi^{(1)} = x_1 - x_2 - \epsilon \,, \qquad
\phi^{(1)} = \xi_3 - \epsilon^{1 + \alpha} \,,
\end{equation}
with positive small parameter $\epsilon\ll 1$ and constant $\alpha \geq 1$.
When $\epsilon$ goes to zero the system tends to the Singular
orbit stratum with two coinciding eigenvalues.
Let us at first find the dynamical consistence of the constraints
(\ref{eq:EpsConst}) considering instead of the Hamiltonian (\ref{eq:hamsma})
the modified Hamiltonian
\footnote{Here we again restrict consideration by the case $\eta^R_a=0$}
\begin{equation} \label{eq:ECMS+cs}
H^\prime  = \frac{1}{8} \sum_{a=1}^3 p_a^2 +
\frac{1}{16} \sum_{(abc)} \xi^2_a \, V(x_b - x_c) +
u \, \left( x_1 - x_2 - \epsilon \right) +
\lambda \, \left( \xi_3 - \epsilon^{1 + \alpha} \right)\,,
\end{equation}
with Lagrangian multipliers $u$ and $\lambda$  and $V(x) = \sinh^{-2}(x)$.
The conservation in time of the constraints $\chi^{(1)}$ and $\phi^{(1)}$
leads to the new  constraints
\begin{equation}\label{eq:CS}
\chi^{(2)} = p_1 - p_2\,, \qquad \phi^{(2)} =  \xi_2 \,,
\end{equation}
while from the maintenance of (\ref{eq:CS}),
the Lagrangian multipliers can be found
\begin{eqnarray}
&& u =
\frac{1}{16} \,
\left[\,
\frac{1}{2}\, \, \xi^2_1 \,\, V^\prime (x_2 - x_3) -
\frac{1}{2} \,\, \xi^2_2 \,\, V^\prime (x_1 - x_3) -
\xi_3^2 \,\, V^\prime (x_1 - x_2)
\right]
\,, \\
&& \lambda =  \frac{1}{8} \,
\xi_3 \,
\left[V(x_2 - x_3) - V(x_1 - x_2)\right]\,.
\end{eqnarray}

Hence we conclude that the constraint surface defined by
(\ref{eq:EpsConst}) and (\ref{eq:CS}) represents
an invariant submanifold for the Hamiltonian (\ref{eq:ECMS+cs}).
Because these  constraints are second class in the
Dirac terminology \cite{BorisovMamaev,DiracL,HenTeit}
we are able to replace the Poisson brackets by the Dirac ones
according to (\ref{eq:Dirbr}) and let the constraint functions to vanish.
One can easy verify that for the canonical variables
$(x, p)$ the corresponding nonzero fundamental Dirac brackets are
\begin{equation}\label{eq:DiracBSOr2}
\{ x_i , p_j \}_D = \frac{1}{2}\,, \quad i, j = 1,2 \,, \qquad
\{x_3, p_3 \}_D = 1\,,
\end{equation}
while for the angular variables we have
\begin{equation}
\{ \xi^R_a , \xi^R_b \}_D = 0 \,,  \qquad a, b = 1,2,3.
\end{equation}
Projecting the Hamiltonian $H^\prime $ to the constraint shell
and then taking the limit $\epsilon \rightarrow 0$
we obtain
\begin{equation}\label{eq:SingOr2}
H^{(2)} : = \lim_{\epsilon \rightarrow 0} H_T\vert_{CS} =
\frac{1}{4}\, p_1^2 + \frac{1}{8} \, p_3^2 +
\frac{{\xi_1}^2}{16 \, \sinh^2(x_1 - x_3)}\,.
\end{equation}
The quantity $\xi^2_1$ in (\ref{eq:SingOr2})
is a constant of motion, that is the reminiscent of
conserved total momentum $\xi_1^2 +\xi_2^2 +\xi_3^2$
for the Euler-Calogero-Moser-Sutherland Hamiltonian (\ref{eq:esham+}).
So, using the appropriate limiting procedure
from the Principle orbit stratum we arrive at
mass deformed Calogero-Moser-Sutherland model corresponding to the Singular stratum,
labeled by two coinciding eigenvalues of $3 \times 3$ symmetric
matrix.
Finally, we establish a relation between (\ref{eq:SingOr2}) and the Hamiltonian
(\ref{eq:HSOr2}) derived before.
To achieve this it is necessary to rescale the momentum
$p_1 \mapsto 2^{- 1} p_1$ so that  variables $(x_1, p_1, x_3, p_3)$
obey canonical Poisson bracket relations instead of
the Dirac brackets (\ref{eq:DiracBSOr2}), identify variables
$x := x_1, \, y := x_3$,  and  constants $ l^2= \xi_1^2 $.  As a result we arrive
at Hamiltonian system which coincides with the mass deformed
II$A_3$ Euler-Calogero-Moser-Sutherland model,
derived in the previous section using explicit parameterizations of
the induced metric on Singular orbit stratum.


\section{Concluding Remarks}


Nowadays we have revival of the interest to matrix models
(see e.g. \cite{Polychronakos}) connected with the search of relations
between the supersymmetric Yang-Mills theory and integrable
systems (for a modern review see \cite{DPReview}).
As it has been shown in the recent paper \cite{ECM-YM} the
Euler-Calogero-Moser-Sutherland model with certain external
potential describes the gauge invariant long-wavelength
approximation of the $SU(2)$ Yang-Mills field theory \cite{YMTheory}.
In the context of the consideration of higher dimensional gauge groups it is
interesting to explore the mechanics on the general linear group
manifold.
In the present Letter we have considered the simplest
version of geodesic motion on $\g$ group manifold and
and analyze the dynamics in the context of isometries
of the bi-invariant metric. Namely we intensively exploit the
slice structure of $\g$
based on the existence of the Principal orbit
stratum and Singular orbit strata of the $\son$ group action.
We demonstrated that the free motion
on the Principal orbit stratum corresponds to the integrable
many-body system of free particles on a line with two types of
internal variables called ``spin'' and ``isospin'', which is a
generalization of the Euler-Calogero-Moser-Sutherland model.
To clarify its relation to the known integrable models we have
implemented two different types of reduction: due to discrete
symmetry and due to continuous symmetry.
In both cases we derived II$A_n$ Euler-Calogero-Moser-Sutherland model.
Concerning the Singular orbit strata, it was shown that in this case the
corresponding dynamical system is the certain deformation of the
Calogero-Moser-Sutherland model in a sense of description of
particles with different masses.
The masses of the particles are not
arbitrary, their mass ratios depend on the degeneracy of the given
Singular orbit stratum.
As example was considered the case of $\gt$ group,
restricted to the Singular orbit stratum with two coinciding eigenvalues.
In this case the reduced system coincides with the II$A_2$
Calogero-Moser-Sutherland model with particle mass ratio
$m_1 = 2 \, m_2 $.

\begin{ack}
It is a pleasure to thank B. Dimitrov, V.I. Inozemtsev, A.N.
Kvinikhidze, M.D. Mateev, and P. Sorba for illuminating discussions.
We are grateful to the Referee for his detailed comments
allowing us to improve the presentation.
The work was supported in part by RFBR  grant
01-01-00708 (AK and DM) and INTAS grant 00-00561 (AK).
\end{ack}



\begin{thebibliography}{99}
\bibitem{CSM}
F. Calogero,
J. Math. Phys. 10 (1969) 2191;
J. Math. Phys. 10 (1969) 2197;
J. Math. Phys. 12 (1971) 419;
B. Sutherland,
Phys. Rev. A 4 (1971) 2019;
Phys. Rev. A 5 (1972) 1372;
J.~Moser,
Adv. Math. 16 (1975) 197.
\bibitem{PerelBook}
A.M. Perelomov,
Integrable Systems of Classical Mechanics and Lie Algebras, Vol. I,
Birkh\"auser, Basel-Boston-Berlin, 1990.
\bibitem{GibbonsHermsen}
J. Gibbons and T. Hermsen,
Physica D 11 (1984) 337.
\bibitem{Wojciechowski}
S. Wojciechowski,
Phys. Lett. A 111 (1985) 101.
\bibitem{KBBT}
I.M. Krichever, O. Babelon, E. Billey, and M. Talon,
Am. Math. Soc. Trasl. (2) 170 (1995) 83.
\bibitem{BabelonTalon}
O. Babelon and M. Talon,
Phys. Lett. A 236 (1997) 462.
\bibitem{LiXu1}
Luen-Chau Li and Ping Xu,
C.R. Acad. Sci. Paris, Serie I, 331 (2000) 55.
\bibitem{LiXu2}
Luen-Chau Li and Ping Xu,
Integrable Spin Calogero-Moser systems,
[arXiv: math.QA/0105162].
\bibitem{Arnold}
V.I. Arnold,
Mathematical Methods of Classical Mechanics,
Springer, Berlin-Heidelberg-New-York, 1978.
\bibitem{MarsdenRatiu}
J. Marsden and T.S. Ratiu,
Introduction to mechanics and symmetry,
Springer, Berlin-Heidelberg-New-York, 1994.
\bibitem{BorisovMamaev}
A.V. Borisov and I.S. Mamaev,
Poisson structures and Lie algebras in Hamiltonian mechanics,
Regular and Chaotic Dyanamics Vol. VII,
University of Udmurtia, Izhevsk, 1999 (in Russian).
\bibitem{OlshanetskyPerelomov}
M.A. Olshanetsky and A.M. Perelomov,
Phys. Rep. 94 (1983) 313.
\bibitem{KazhdanKonstantSternberg}
D. Kazhdan, B. Konstant and S. Sternberg,
Comm. Pure. Appl. Math. 31 (1978) 481.
\bibitem{GorskiNekrasov}
A. Gorsky and N. Nekrasov,
Elliptic Calogero-Moser system from two-dimensional current algebra,
[arXiv: hep-th/9401021].
\bibitem{PolyDiscrete}
A.P. Polychronakos, Nucl. Phys. B543 (1999) 485.
\bibitem{Veselov}
A.P. Veselov,
in "Calogero-Moser-Sutherland models",
Ed. Jan Felipe van Diejen and Luc Vinet,
CRM Series in Mathematical Physics,
Springer, Berlin, New-York, 2000.
\bibitem{Zelobenko}
D.R. Zelobenko,
Compact Lie groups and their representations,
Translations of Mathematical Monographs, Vol. 40 AMS, 1978.
\bibitem{DPLaxRep}
E. D'Hoker and D.H. Phong,
Lax pairs and spectral curves for Calogero-Moser and spin Calogero-Moser systems,
[arXiv: hep-th/9903002].
\bibitem{DiracL}
P.A.M. Dirac,
Lectures on Quantum Mechanics,
Belfer Graduate School of Science, Yeshiva University Press, New York, 1964.
\bibitem{HenTeit}
M. Henneaux and C. Teitelboim,
Quantization of Gauge Systems,
Princeton University Press, Princeton, NJ, 1992.
\bibitem{Mehta}
M.L. Mehta,
Random matrices,
Boston, MA: Academic, 1991.
\bibitem{Polychronakos}
A.P. Polychronakos,
Generalized statistics in one dimension,
[arXiv: hep-th/9902157].
\bibitem{DPReview}
E. D'Hoker and D.H. Phong,
Lectures on supersymmetric Yang-Mills theory and integrable systems,
[arXiv: hep-th/9912271].
\bibitem{ECM-YM}
A.M. Khvedelidze and D.M. Mladenov,
Phys. Rev. D62 125016 (2000).
\bibitem{YMTheory}
S.A. Gogilidze, A.M. Khvedelidze, D.M. Mladenov, and H.-P.Pavel,
Phys. Rev. {\bf D57}, (1998) 7488;
A.M. Khvedelidze and H.-P. Pavel,
Phys. Rev. {\bf D 59} (1999) 105017;
A.M. Khvedelidze, D.M. Mladenov, H.-P. Pavel and G. R\"opke,
Unconstrained $SU(2)$ Yang-Mills theory with topological term
in the long-wavelength approximation,
[arXiv:hep-th/02022145].
\end{thebibliography}
\end{document}